\documentclass[iop]{emulateapj-rtx4}

\usepackage{graphicx}
\usepackage{amssymb}
\usepackage{amsmath}
\usepackage{natbib}
\nocite{*}

\submitted{Accepted to The Astronomical Journal July 28, 2014}

\shortauthors{TOFFLEMIRE ET AL.}
\shorttitle{WIYN OPEN CLUSTER STUDY. LIX}
\begin{document}

\title{WIYN Open Cluster Study. LIX. Radial-Velocity Membership of the Evolved
  Population of the Old Open Cluster NGC 6791}  

\author{Benjamin M. Tofflemire,\altaffilmark{1, 2}
Natalie M. Gosnell,\altaffilmark{1, 2}
Robert D. Mathieu,\altaffilmark{1, 2}
and Imants Platais\altaffilmark{3}
}

\altaffiltext{1}{Department of Astronomy, University of Wisconsin--Madison,
  475 North Charter Street, Madison, WI 53706, USA; tofflemi@astro.wisc.edu}
\altaffiltext{2}{Visiting Astronomer, Kitt Peak National Observatory, National
  Optical Astronomy Observatory, which is operated by the Association of
  Universities for Research in Astronomy (AURA) under cooperative agreement
  with the National Science Foundation.}  
\altaffiltext{3}{Department of Physics and Astronomy, The Johns Hopkins
  University, Baltimore, MD 21218, USA; imants@pha.jhu.edu}

\begin{abstract}
  The open cluster NGC 6791 has been the focus of much recent study due to its
  intriguing combination of old age and high metallicity ($\sim$8 Gyr,
  [Fe$/$H]=$+$0.30), as well as its location within the {\sl Kepler} field. As
  part of the WIYN Open Cluster Study, we present precise ($\sigma=0.38$ km
  s$^{-1}$) radial velocities for proper-motion candidate members of NGC 6791
  from Platais et al. Our survey, extending down to $g^\prime\sim16.8$, is
  comprised of the evolved cluster population, including blue stragglers,
  giants, and horizontal branch stars. Of the 280 proper-motion-selected stars
  above our magnitude limit, 93\% have at least one radial-velocity
  measurement and 79\% have three measurements over the course of at least 200
  days, sufficient for secure radial-velocity-determined membership of
  non-velocity-variable stars. The Platais et al. proper-motion catalog
  includes twelve anomalous horizontal branch candidates blueward of the red
  clump, of which we find only four to be cluster members. Three fall slightly
  blueward of the red clump and the fourth is consistent with being a blue
  straggler. The cleaned color-magnitude diagram shows a richly populated red
  giant branch and a blue straggler population. Half of the blue stragglers
  are in binaries. From our radial-velocity measurement distribution we find
  the cluster's radial-velocity dispersion to be $\sigma_c=0.62\pm0.10$ km
  s$^{-1}$. This corresponds to a dynamical mass of $\sim$4600 $M_\odot$.
\end{abstract}


\section{INTRODUCTION}
\label{intro}

Galactic open star clusters are a key laboratory for observationally
constraining models of stellar evolution. The open cluster NGC 6791 resides in
a sparsely populated area of the age and metallicity parameter space with its
unique combination of old age ($\sim$8 Gyr; \citealt{Grundahletal2008}) and
high metallicity ([Fe/H]=+0.30; \citealt{Boesgaardetal2009}). As such, it has
been the subject of many photometric surveys
\citep{Kinman1965,Kaluzny&Rucinski1995,Stetsonetal2003,Plataisetal2011,
  Carraroetal2013}, spectroscopic studies
\citep{Liebertetal1994,Carraroetal2006,Boesgaardetal2009,Frinchaboyetal2013},
eclipsing binary studies \citep{Grundahletal2008,Brogaardetal2012}, variable
star surveys \citep{Hartmanetal2005,Mochejskaetal2005,deMarchietal2007}, and
an X-ray binary survey \citep{vandenBergetal2013}. Falling within the {\sl
  Kepler} field \citep{Boruckietal2010}, NGC 6791 also has asteroseismic
measurements of select red giant (RG) and red clump (RC) stars
\citep{Stelloetal2011,Corsaroetal2012}.

With this wealth of data, NGC 6791 has proven difficult to describe in terms
of a single-age stellar population. Evidence for extended star formation
\citep{Twarogetal2011}, blue straggler stars \citep[BSSs;][]{Kinman1965}, red
stragglers \citep[sub-subgiants;][]{Plataisetal2011}, young white dwarfs
\citep{Bedinetal2005,Kaliraietal2007}, and extreme horizontal branch (EHB) and
blue-horizontal branch stars in addition to a rich RC
\citep{Kaluzny&Udalski1992,Greenetal1996} make this open cluster a popular
target of observers and theorists alike. The potential presence of EHB stars,
for instance, may make NGC 6791 our closest analog in studying the ultraviolet
upturn of elliptical galaxies \citep{Buzzonietal2012}. With a diversity of
stellar sources for study, secure membership determinations remain a foremost
priority before extensive future analyses.

The \citet[hereafter P11]{Plataisetal2011} proper-motion (PM) study provides
the first comprehensive, kinematic membership determination for NGC 6791. A
previous PM study by \citet{Cudworth&Anthony-Twarog1993} was limited to only
the inner $3\arcmin$. Other memberships based on radial-velocity (RV)
measurements, eclipsing binaries, spectroscopically derived stellar
parameters, and/or asteroseismic measurements have, thus far, been limited to
small sample sizes.

A surprising finding of the P11 study was a proposed population of 12
horizontal branch stars (7 having proper-motion membership probabilities above
90\%) extending blueward continuously from the RC into the BSS domain
(hereafter referred to as blue-HB candidates). Given the high metallicity of
the cluster ([Fe/H]$ = + 0.30$), core helium-burning stars are expected to
reside within the RC, making this seeming horizontal branch an
anomaly. \citet{Brogaardetal2012} point out that in a Johnson $BV$
color-magnitude diagram (CMD) a model
zero-age horizontal branch is more luminous than the locus of blue-HB stars
proposed by P11. Therefore, these authors suggest that the blue-HB candidates
must instead be BSSs. Inconsistencies like this motivate our investigation
into the nature of these stars using multi-epoch, high-resolution
spectroscopy.

\clearpage

Through the WIYN\footnote{The WIYN Observatory is a joint facility of the
  University of Wisconsin-Madison, Indiana University, Yale University, and
  the National Optical Astronomy Observatory.} Open Cluster Study
\citep[WOCS;][]{Mathieu2000}, we provide the first, high-precision, systematic
RV survey of the evolved population of NGC 6791 within $30\arcmin$ of the
cluster center. Combining high-precision RV and PM memberships, we define
secure, three-dimensional kinematic memberships for a complete stellar sample
including RGs, HB stars, and BSSs. Our ongoing survey allows us to also
characterize the binary population and determine the cluster's radial-velocity
dispersion from which we derive the first dynamical mass estimate for this
cluster. Our paper is structured as follows: Section \ref{sample} describes
our survey sample, Section \ref{obs+dr} details our observation and data
reduction scheme, and Section \ref{comp} presents our observation
completeness. Sections \ref{results} and \ref{disc} present our results and
discussion, respectively, and a summary of our conclusions is provided in
Section \ref{conc}.

\section{STELLAR SAMPLE}
\label{sample}

The source catalog for our RV survey of NGC 6791 is drawn from the PM study of
P11. In this study, deep photographic plates dating back to 1961 from the Kitt
Peak National Observatory (KPNO) 4 m and the Lick 3 m telescopes, along with
CCD mosaic frames taken from 1998--2009 with the KPNO 4 m and the 3.6 m
Canada-France-Hawaii Telescope (CFHT), were used to obtain high-precision
proper motions for 58,901 stars down to $g^\prime \sim 24$. Within the 0.8
deg$^2$ field of view (FOV), a total of 5,699 PM probable cluster members were
found based on the chosen cluster membership probabilities combined with
certain CMD restrictions (Figure 1 in P11). In general, stars with PM
membership probabilities $P_{\mu} \ge 19\%$ were taken to be probable cluster
members, although a looser constraint of $P_\mu \ge 1\%$ was used for stars
falling on the main sequence. The complete catalog of positions, proper
motions, membership probabilities, and $g^\prime$ and $r^\prime$ photometry
will appear in I. Platais et al. (2014, in preparation). 

Our RV catalog is drawn from all P11 sources above a limiting magnitude of
$g^\prime \le 16.8$, set by the minimum signal-to-noise ratio required by our
survey. They comprise the evolved stellar population including the RG branch
(RGB), RC, possible blue-HB stars, and BSSs. We trim the full sample only by
excluding $P_\mu = 0\%$ sources, assuming they are field stars. This
assumption is supported by the CMD of $P_\mu = 0\%$ stars, presented in Figure
1, which lacks apparent cluster features such as the RGB. The CMD membership
cut of $P_\mu \ge 19\%$ used by P11, aimed to minimize the field contamination
expected in low-$P_\mu$ sources. In the $1\% \le P_\mu < 19\%$ sample,
however, some cluster members are indeed expected. Thus, we include them in our
RV catalog with the expectation that RV membership information will draw out
the cluster members. Applying this selection criterion ($P_\mu > 0\%$) yields
a total sample to 280 sources.

\begin{figure}[!tbh]
  \centering
    \includegraphics[keepaspectratio=true,scale=0.45]{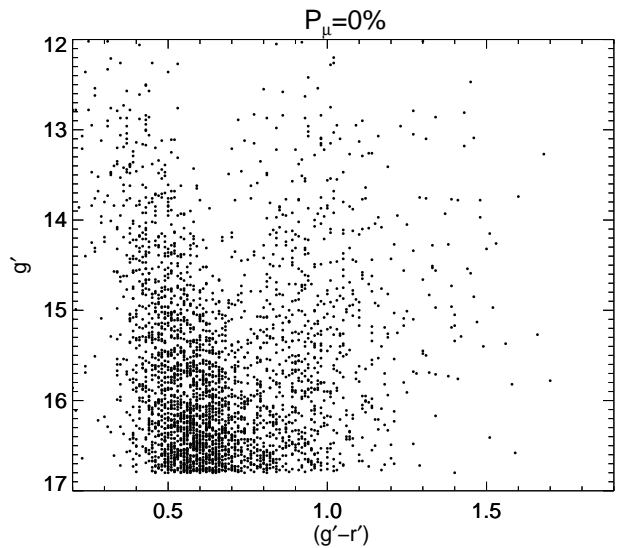}
    \caption{CMD of proper-motion-defined field stars
      ($P_\mu=0\%$). Signatures of cluster members, like the RGB, are
      absent. (Compare with Figure \ref{fig:zcmd} below.)}
    \label{fig:pmzero}
\end{figure}

\section{OBSERVATIONS \& DATA REDUCTION}
\label{obs+dr}

Here we briefly present our observing and data reduction procedures. (We refer
readers to \citet{Gelleretal2008} for full details.) Variations in our methods
from previous WOCS RV papers are given below.

\subsection{Spectroscopic Observations}
\label{obs}

Our ongoing spectroscopic survey of NGC 6791 has, to date, yielded 1189 RV
measurements of 260 stars. Observations utilize the Hydra Multi-Object
Spectrograph (MOS) on the WIYN 3.5 m telescope, capable of obtaining $\sim$70
stellar spectra simultaneously over a 1$^{\circ}$ field of view. We use $3
\farcs 1$ diameter fibers along with the bench spectrograph echelle grating,
resulting in a spectral resolution of $\sim$20,000 (15 km s$^{-1}$). Due to
the number of source fibers, the echelle grating is not cross-dispersed and
the X14 filter is used to isolate the 11th spectral order. This 250 \AA-wide
region centered at 5125 \AA \ contains many stellar absorption lines including
the Mg I b triplet.

To begin our observing program, the catalog is split into bright
($12<g'<15.9$) and faint ($12.9<g'<16.8$) sub-samples to prevent on-chip
contamination between source spectra. Bright and faint Hydra configurations
generated from these sub-samples are observed for 1 and 2 hours, respectively,
to ensure a sufficient signal-to-noise ratio (S/N $\simeq 18$ per resolution
element under ideal observing conditions for all targets). Integrations are
split into three equal exposures to reduce the impact of cosmic ray
contamination. Of the 81 available fibers, each configuration observes a
maximum of 71 stellar spectra with 10 fibers reserved for sky sampling. As
accurate wavelength calibration is vital to this program's success, comparison
ThAr emission lamp spectra are taken before and after each configuration to
correct for wavelength shifts during the observing sequence. Dome flat-field
images are also taken for each configuration to aid in throughput correction
and aperture identification during spectral extraction.

Target prioritization aims to efficiently determine the RV membership of
proper-motion-selected stars. Monte Carlo (MC) simulations have shown that 3
observations over the course of 200 days can determine, to $\sim$90\%
confidence, whether a star has a constant or variable RV for binary orbital
periods up to 1000 days \citep[following][]{Geller&Mathieu2012}. As such, for
a given observing run, we place stars with one observation at the highest
priority followed by twice-observed likely members (stars whose first two RV
measurements fall within the cluster's RV distribution), twice-observed likely
non-members, unobserved stars, and finally, stars with three or more
observations over the course of at least 200 days. The 12 blue-HB candidates
from P11 were given the highest priority until their RV memberships were
determined. Unlike previous WOCS RV studies, we have not yet chosen to
prioritize binary orbit determinations for velocity-variable stars in favor of
maximizing the number of single stars with determined RV memberships. Stars
with the highest PM probabilities were also prioritized in our observations.

\subsection{Data Reduction}
\label{dr}

All WOCS image processing and data reduction are completed using standard
IRAF\footnote{IRAF is distributed by the National Optical Astronomy
  Observatory, which is operated by the Association of Universities for
  Research in Astronomy (AURA) under cooperative agreement with the National
  Science Foundation.} tasks. Our routine is as follows. Bias subtraction from
an overscan region is first applied to raw science and flat-field
images. Spectral extraction, flat-field correction, throughput correction, and
dispersion solutions are computed and applied with the {\sl dohydra}
reduction task. Remaining sky fibers are visually inspected for quality
and median combined to perform sky subtraction. The three fully reduced
spectra from each configuration are then median combined to improve
S/N and remove cosmic ray contamination.

Fourier cross-correlation functions (CCF) between stellar spectra and a
zero-velocity template are computed using the IRAF task {\sl fxcor}. A
Gaussian fit to the CCF peak provides the relative velocity shift between the
two spectra, which is then corrected for the Earth's orbital motion to produce
a heliocentric RV. \citet{Gelleretal2008} define a minimum CCF peak height of
0.4 as sufficient correlation for a secure RV measurement utilizing the same
observational and data reduction schemes. We adopt this same minimum CCF peak
value to identify acceptable RV measurements. Finally, systematic
fiber-to-fiber RV offsets derived by \citet{Gelleretal2008} are then applied
to the computed heliocentric RV, yielding our final RV measurement.

In the case of double-lined spectroscopic binaries, we are able to retrieve
the radial velocities of both stars simultaneously using the two-dimensional
correlation routine TODCOR \citep{Zucker&Mazeh1994}. Using a zero-velocity
template for each star, we fit a two-dimensional correlation that is able to
effectively decompose strongly blended lines. The result of this routine is a
heliocentric RV for each star in the system that is then corrected for
fiber-fiber variations.

As in previous WOCS surveys, the standard Fourier cross-correlation template
is a high-S/N solar spectrum from a sky flat. For most stars in the NGC 6791
sample, a strong correlation peak is found using the solar template. However, a
handful of RGs in our sample exhibit the Swan C$_{2}$ absorption band near
5165\AA \ that prevents precise RV measurements against a solar template. For
this subset, we use the spectrum of the K1 III star, HD172171, from the
publicly available ELODIE Stellar Library \citep{Prugniel&Soubiran2001} as our
template. This spectrum is provided at zero-velocity and is set to a
zero-heliocentric-velocity for cross-correlation. Containing the same C$_{2}$
absorption feature and at a spectral resolution of $\sim$42,000, we retrieve
strong correlation peaks for this subset of RGs in our sample. Additionally,
the RV zero-points for cluster members found using the solar and RG template
are consistent.

\begin{figure}[!t]
  \centering
  \includegraphics[keepaspectratio=true,scale=0.45]{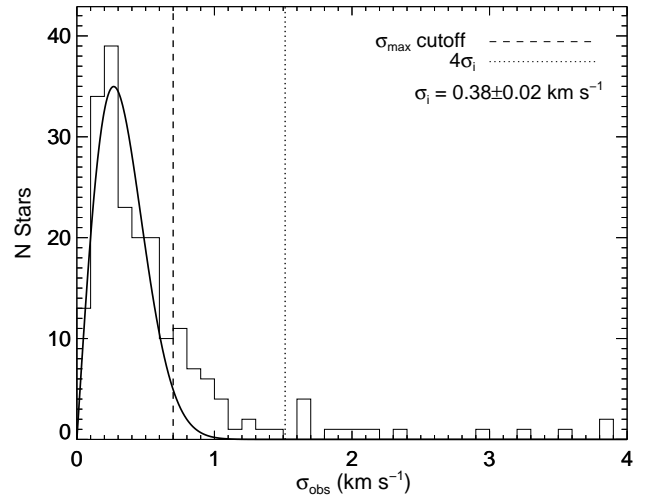}
    \caption{Distribution of the standard deviation of the first 3 RV
      measurements for each star. The solid curve displays the best fit of the
      $\chi^2$ model within the $\sigma_{\rm max}$ cutoff represented by the
      dashed vertical line. A best-fit precision of $\sigma_i=0.38\pm0.02$ km
      s$^{-1}$ is found. The vertical dotted line marks 4$\sigma_i$, the
      minimum standard deviation required for a velocity-variable
      designation.}
    \label{fig:precision}
\end{figure}

\subsection{Precision}
\label{precision}

To assess our measurement precision we fit the standard deviations of RV
measurements with a $\chi^2$ distribution following
\citet{Gelleretal2008}. For consistency across our sample, we take only
objects with at least three observations, and use only the first three RV
measurements when calculating standard deviations ($\sigma_{\rm obs}$). The
$\chi^2$ function models the distribution of error for measurements of single
(non-velocity-variable) stars. We fit the following form,
\begin{equation}
    \chi^2(\sigma_{\rm obs})=
        A\left(\frac{2}{\sigma^2_i}\right)
        (\sigma_{\rm obs}) {\rm exp} 
        \left(-\frac{\sigma^2_{\rm obs}}{\sigma^2_i}\right),
\end{equation}
with two free parameters: $A$, the $\chi^2$ normalization, and $\sigma_i$, our
RV measurement precision. The velocity variability of binary stars in our
sample inflates the tail of our error distribution with sources of high
measurement deviation. To account for this, we fit $\chi^2$ only to data
within a $\sigma_{\rm max}$ cutoff to limit binary contamination. Varying
$\sigma_{\rm max}$ at 0.1 km s$^{-1}$ intervals we find that a value of
$\sigma_{\rm max}=0.7$ km s$^{-1}$ produces the lowest residuals in the fit
and we adopt this value as our cutoff. 

\begin{figure*}[!tbh]
\centering
\includegraphics[keepaspectratio=true,scale=0.44]{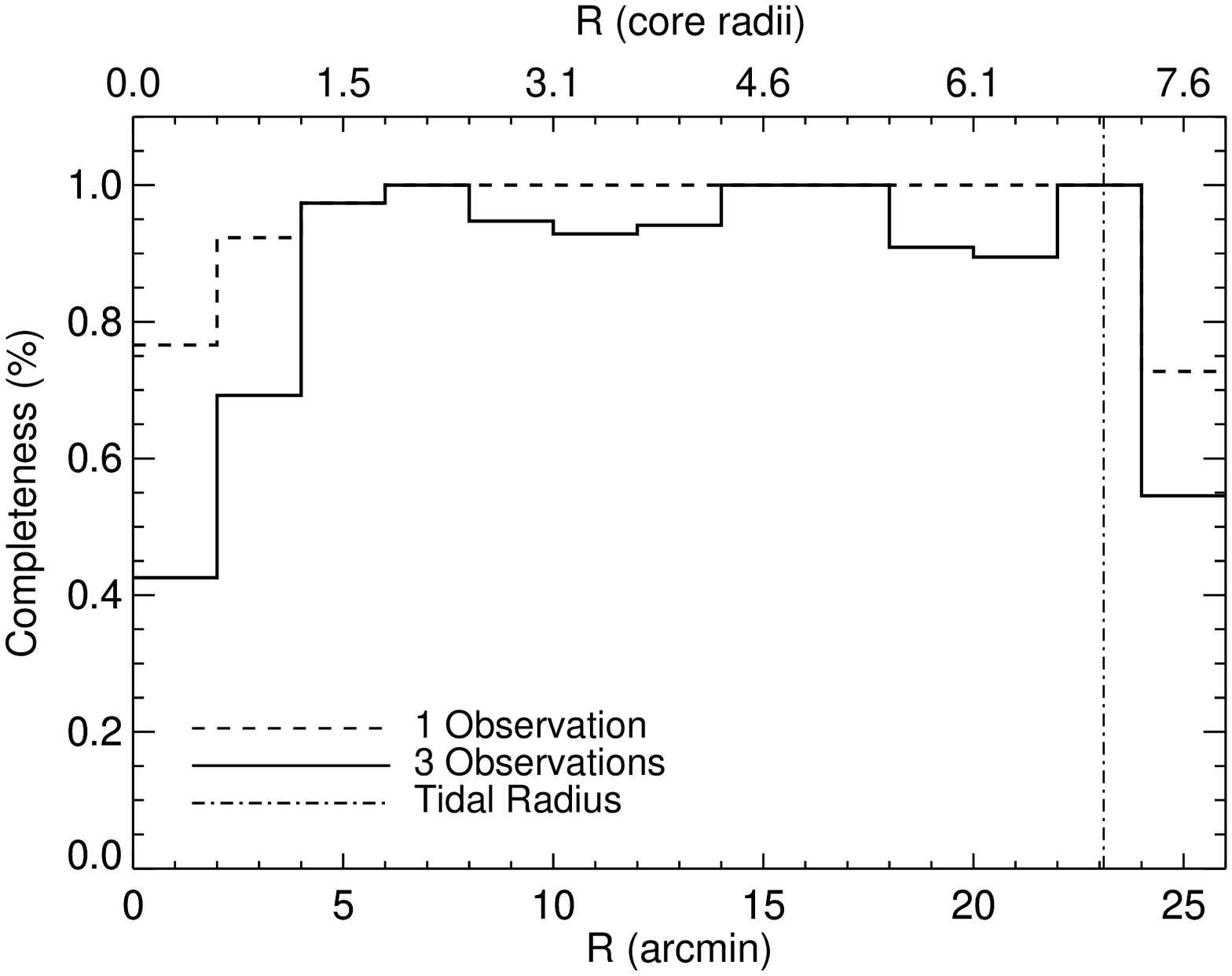}
\includegraphics[keepaspectratio=true,scale=0.44]{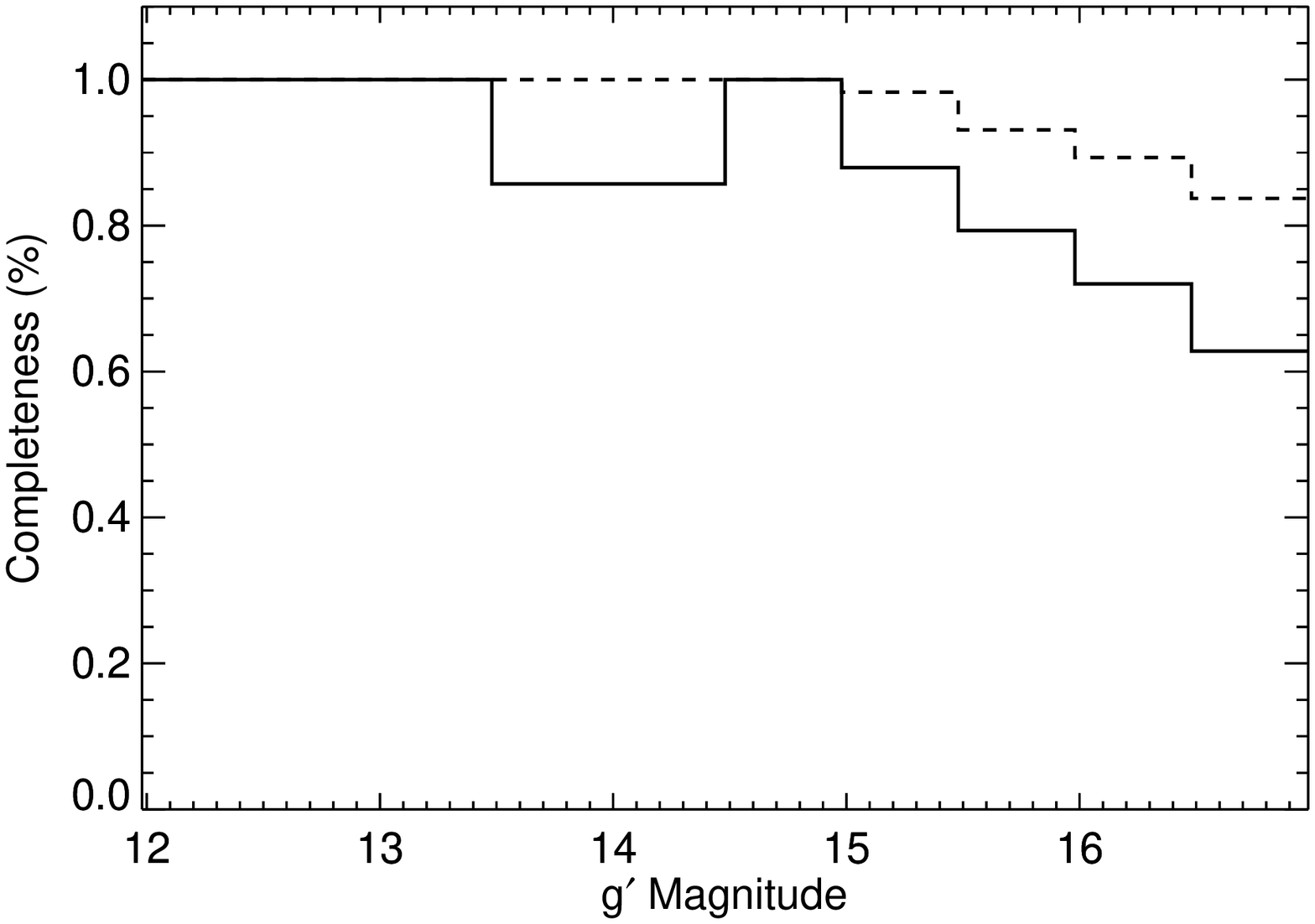}
\caption{Completeness histograms as a function of radius and $g^\prime$
  magnitude for our sample ($P_\mu \ge 1\%$). Dashed lines show the
  completeness of stars with $\ge 1$ radial-velocity measurements. Solid lines
  show the completeness of stars with $\ge 3$ measurements over the course of
  at least 200 days (for which we can determine radial-velocity
  membership). In the left panel we present radius in both arcminutes and core
  radii for R$_c=3.8$ pc (P11) at a distance of 4 kpc
  \citep{Grundahletal2008}. The dot-dashed vertical line marks the tidal
  radius of the cluster at $23\farcm1$ ($\sim$27 pc; P11).}
    \label{fig:comp}
\end{figure*}

Figure \ref{fig:precision} presents our empirical RV standard deviations with
the best-fit error distribution over-plotted as a solid line. The dashed
vertical line signifies the $\sigma_{\rm max}$ cutoff. Contamination from
velocity variables can be clearly seen exceeding the $\chi^2$ fit beyond the
cutoff. We find a best-fit precision of $\sigma_i=0.38\pm0.02$ km s$^{-1}$
that is insensitive to the value of $\sigma_{\rm max}$ chosen, producing a
consistent result for $\sigma_{\rm max}=0.6-0.8$ km s$^{-1}$. Our best-fit
measurement precision is consistent with that found in the
\citet{Gelleretal2008} study of NGC 188 ($\sigma_i=0.4$ km s$^{-1}$).

From our RV precision, we define a minimum standard deviation in observed RV
for a star to be characterized as a ``velocity-variable''. As in
\citet{Gelleretal2008}, we use the quantity $e/i$ to define velocity variables
where $e$ is the observed standard deviation of a given star ($\sigma_{\rm
  obs}$) and $i$ denotes our precision ($\sigma_i$). Stars with $e/i > 4$ are
designated as velocity variables. The $e/i$ limit is plotted as the vertical
dotted line in Figure \ref{fig:precision}, where stars falling right of this
line are considered to be velocity variable.

\section{COMPLETENESS}
\label{comp}

By combining precise RV and PM memberships, we seek to define a complete
sample of the evolved stellar population within a radius of $30\arcmin$
($\sim$35 pc at $d=4$ kpc) from the cluster center. We adopt the
\citet{Stetsonetal2003} cluster center in the following analysis ($\alpha =
19^{\rm h}20^{\rm m}53^{\rm s}; \delta = +37^\circ46\arcmin30\arcsec$; J2000),
which is within $2\arcsec$ from our independent estimate. Within the magnitude
range of our sample (see Section 2), there are 280 sources ($P_\mu \ge
1\%$). Of these, 93\% have at least one RV measurement and 79\% have $\ge$ 3
measurements over the course of at least 200 days. As stated above, we have
presumed that stars with $P_\mu = 0\%$ are securely not cluster members.

Achieving 3 RV measurements over the course of at least 200 days is required
to determine, with $\sim$90\% confidence, whether a star is velocity
variable. Since RV membership probabilities are only computed for
non-velocity-variable stars (Section \ref{membership}), this observational
benchmark sets the completeness for which we can determine membership or
binarity.

Distributions of RV measurement completeness with respect to radius and
$g^\prime$ magnitude are presented in the left and right panels of Figure
\ref{fig:comp}, respectively. The dashed line represents completeness with
$\ge 1$ measurement. Completeness with $\ge$ 3 measurements over 200 days is
shown as the solid line. Radius is given in arcminutes (bottom) and in core
radii (top) based on the P11 derivation of $R_c\simeq3.8$ pc from King model
fitting \citep{King1966} to the stellar number-density profile (assuming a
distance of 4 kpc, 1\arcsec=1.16pc; \citealt{Grundahletal2008}). The cluster
tidal radius ($\sim$27 pc; P11) is also presented as the vertical dot-dashed
line.

The decline in RV measurement completeness at small radii and at faint
magnitudes is due to the same population of dim, centrally concentrated
stars. Observations of these stars are limited by the number of Hydra fibers
that can be placed in the dense cluster center, compounded with the time
expense of obtaining 2 hr exposures for faint stars.

\begin{deluxetable*}{rrrrrrrrcrcc}[!th]
\tablewidth{0pt}
\tabletypesize{\footnotesize}
\tablecaption{Radial Velocity Summary Table}
\tablehead{
  \colhead{ID$_{\rm W}$} &
  \colhead{R.A.} &
  \colhead{Dec.} &
  \colhead{$g^\prime$} &
  \colhead{$g^\prime$-$r^\prime$} &
  \colhead{N$_{\rm obs}$} &
  \colhead{$\overline{\rm RV}$} &
  \colhead{$P_{\mu}$} &
  \colhead{$P_{\rm{RV}}$\,\tablenotemark{a}} &
  \colhead{$e/i$} &
  \colhead{Class} &
  \colhead{Comment}
}
\startdata
1002 & 
19:20:55.11 & 37:47:16.3 & 14.63 & 1.41 & 5 & -47.51 & 99 & 96 & 0.31 & SM &  \\
1003 & 
19:20:50.04 & 37:47:28.2 & 13.50 & 0.95 & 1 & -49.49 & 10 & ... & ... & U &  \\
1004 & 
19:20:58.63 & 37:47:40.5 & 11.99 & 0.88 & 3 & -13.21 & 10 & 0 & 0.70 & SN &  \\
2003 & 
19:20:47.66 & 37:47:32.3 & 15.33 & 1.23 & 3 & -54.20 & 99 & (0) & 53.30 & BU & SB1 \\
3006 & 
19:20:49.72 & 37:43:42.7 & 14.83 & 1.50 & 4 & -46.47 & 99 & 94 & 1.30 & SM & C$_2$ Band \\
4003 & 
19:20:59.95 & 37:46:03.3 & 15.15 & 0.28 & 13 & -44.76 & 99 & 54 & 1.62 & SM & BSS \\
6006 & 
19:20:49.65 & 37:44:07.8 & 15.18 & 1.10 & 3 & -49.03 & 10 & (89) & 4.26 & BLM & SB1 \\
7021 & 
19:20:33.03 & 37:55:55.9 & 14.21 & 1.10 & 14 & -75.11 & 41 & (0) & 11.00 & BLN & SB1 \\
7045 & 
19:21:22.34 & 38:07:57.1 & 14.04 & 0.39 & 10 & -1.88 & 37 & (0) & 3.22 & SN & RR (113.1 km s$^{-1}$) \\

43033 & 
19:22:15.20 & 37:48:47.2 & 15.82 & 0.55 & 6 & -41.70 & 6 & (0) & 79.28 & BU & SB2 \\

\enddata
\tablecomments{This table is available in its entirety in a machine-readable
  form in the online journal. A portion is shown here for guidance regarding
  its form and contents. Photometry, coordinates, and proper-motion membership
  probabilities come from I. Platais et al (2014, in preparation).}
\tablenotetext{a}{RV membership probabilities in parenthesis indicate the
  probability of the mean RV for velocity variables ($P_{\overline{\rm RV}}$)
  and remains uncertain until a binary orbital solution is found.}
\label{tab:summary}
\end{deluxetable*}

\section{RESULTS}
\label{results}

\subsection{Radial-Velocity Measurements}
\label{rvmeas}

The results of our RV survey are summarized in Table \ref{tab:summary}, where
we present 10 lines selected to illustrate the table format and relevant
comment and class fields. The full table is available electronically. For each
star, we include the WOCS ID (ID$_{\rm W}$), R.A. and Declination (J2000),
$g^\prime$ magnitude, and $(g^\prime$-$r^\prime$) color from the P11
catalog. The succeeding columns present the number of WIYN RV measurements
(N$_{\rm obs}$), mean RV ($\overline{\rm RV}$), PM membership probability
($P_\mu$), RV membership probability ($P_{\rm RV}$), $e/i$ value, membership
class (detailed in Section \ref{membership}), and comments. Both the
$\overline{\rm RV}$ and $e/i$ value are calculated using an error-weighted
mean. RV membership probabilities in parenthesis indicate a tentative value
for velocity variable stars based on their mean RV ($P_{\overline{\rm RV}}$),
which remains uncertain until a binary orbital solution is found. In the
comment field, we highlight the following stars: velocity variables as single-
or double-lined spectroscopic binaries (``SB1'' and ``SB2'', respectively),
RGs containing strong C$_2$ absorption for which a KIII stellar template was
used (``C$_2$ Band''), BSSs (``BSS''), and finally, rapidly rotating stars
(``RR''), defined as having a CCF FWHM $\ge 60$ km s$^{-1}$, are presented
with their projected $v$sin $i$ rotational velocity in parenthesis.

Stars designated as RR have reduced RV measurement precision due to their
broadened photospheric absorption lines and correspondingly broad CCF
peaks. The \citet{Gelleretal2010} study of the young open cluster M35 found
that RV measurement precision scaled linearly with the projected $v$sin $i$
rotational velocity and had the functional form
$\sigma_i=0.38+0.012(v\rm{sin}\ i)$ km s$^{-1}$. We find that 10 stars in our
sample ($\sim4\%$) are defined as RRs and following \citet{Gelleretal2010}, we
derive $v$sin $i$ for these stars from their CCF FWHM. The largest $v$sin $i$
found is $\sim$110 km s$^{-1}$. For these stars, the $e/i$ value and
velocity-variable designation are computed based on their $v$sin $i$ dependent
$\sigma_i$ value. None of the 10 RRs are RGB or RC stars as expected; however,
3 are binary BSS candidates (see Section \ref{bs}).

Table \ref{tab:data} presents every NGC 6791 RV measurement and CCF peak
height obtained with the WIYN telescope through 2014 February. We have
included only a selection of the table here for brevity. The full table is
available electronically. The right two columns are reserved for SB2s when a
velocity and CCF peak height for each star can be obtained.

\

\subsection{Radial-Velocity Membership}
\label{membership}

\begin{figure}[!t]
\centering
\includegraphics[keepaspectratio=true,scale=0.45]{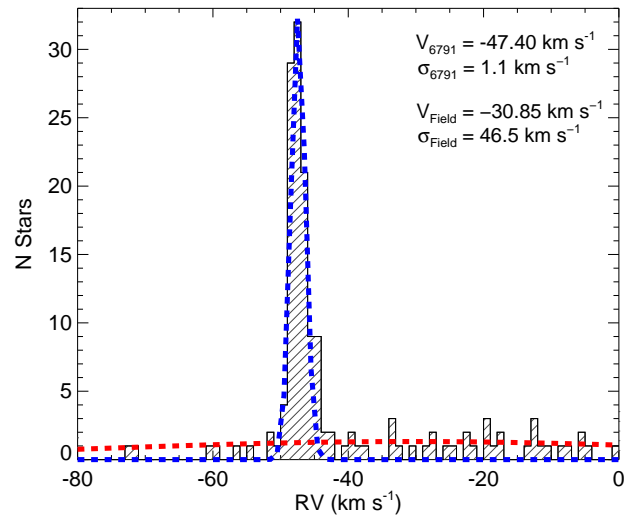}
\caption{Radial-velocity histogram of single, non-rapidly rotating
  stars. Gaussian fits to the field and cluster RV distributions are
  over-plotted in red and blue respectively. Top right provides the
  radial-velocity mean and standard deviation for each of these fits.}
    \label{fig:rvhist}
\end{figure}

Calculations of RV membership require a decomposition of the RV distribution
into field-star and cluster distributions. To separate these two components,
we fit our distribution of RV measurements for the entire sample ($P_\mu \ge
1\%$) simultaneously with two Gaussians. Figure \ref{fig:rvhist} presents our
RV measurement histogram of single, non-rapidly rotating stars with the
best-fit field-star and cluster population models over-plotted in red and
blue, respectively. The fit is made to an 80 km s$^{-1}$ region centered
around the cluster peak. From this procedure, we find a cluster RV of
$-47.40\pm0.13$ km s$^{-1}$ and RV standard deviation of
$\sigma_{6791}=1.1\pm0.1$ km s$^{-1}$. The true one-dimensional velocity
dispersion of the cluster, $\sigma_c$, is less than the Gaussian fit
$\sigma_{6791}$ due to inflation from binary contamination and our measurement
precision. These issues are addressed in Section \ref{mass}. 

\ 

\

\

With a model of the two populations, we define RV membership with the
following equation:
\begin{equation}
    P_{\rm RV}(v)=\frac{F_{\rm cluster}(v)}{F_{\rm field}(v) + 
        F_{\rm cluster}(v)}, 
    \label{eqn:prob}
\end{equation}
where $F_{\rm cluster}(v)$ is the cluster-model value for the stellar RV, $v$,
and $F_{\rm field}(v)$ is the field-star-model value for the same RV
\citep{Vasilevskisetal1958}.  We find the field
distribution to be well separated from the field velocity center ($-30.85$ km
s$^{-1}$), providing a good separation in membership probability between
cluster and field stars.

\begin{deluxetable*}{rcrcrc}[!th]
\tablewidth{0pt}
\tabletypesize{\footnotesize}
\tablecaption{Radial Velocity Data Table}
\tablehead{
  \colhead{ID$_{\rm W}$} &
  \colhead{HJD-2,450,000} &
  \colhead{RV$_1$} &
  \colhead{Correlation} &
  \colhead{RV$_2$} &
  \colhead{Correlation} \\
  \colhead{} &
  \colhead{(days)} &
  \colhead{(km s$^{-1}$)} &
  \colhead{Height$_1$} &
  \colhead{(km s$^{-1}$)} &
  \colhead{Height$_2$} 
}
\startdata
1002 & & & & & \\
 & 5754.679 & -47.5 & 0.63 & & \\
 & 5821.717 & -47.4 & 0.60 & & \\
 & 5995.996 & -47.5 & 0.61 & & \\
 & 6053.821 & -47.4 & 0.60 & & \\
 & 6524.863 & -47.7 & 0.62 & & \\
1003 & & & & & \\
 & 6524.863 & -49.5 & 0.93 & & \\
1004 & & & & & \\
 & 5754.751 & -12.9 & 0.95 & & \\
 & 5821.717 & -13.0 & 0.95 & & \\
 & 6705.980 & -13.4 & 0.95 & & \\
43033 & & & & & \\
 & 5755.698 & -31.4 & 0.75 & & \\
 & 5821.717 & -27.1 & 0.89 & & \\
 & 5995.996 & -2.9 & 0.67 & -60.9 & 0.57 \\
 & 6052.802 & -57.8 & 0.72 & -1.1 & 0.66 \\
 & 6524.863 & -49.5 & 0.66 & -12.4 & 0.51 \\
 & 6705.980 & -8.0 & 0.58 & -55.0 & 0.50 \\
\enddata
\tablecomments{The two right columns are reserved for double-lined
  spectroscopic binaries when a velocity and CCF peak height can be obtained
  for each star. This table is available in its entirety in a machine-readable
  form in the online journal. A portion is shown here for guidance regarding
  its form and contents.}
\label{tab:data}
\end{deluxetable*}

\begin{figure}[!t]
\centering
\includegraphics[keepaspectratio=true,scale=0.45]{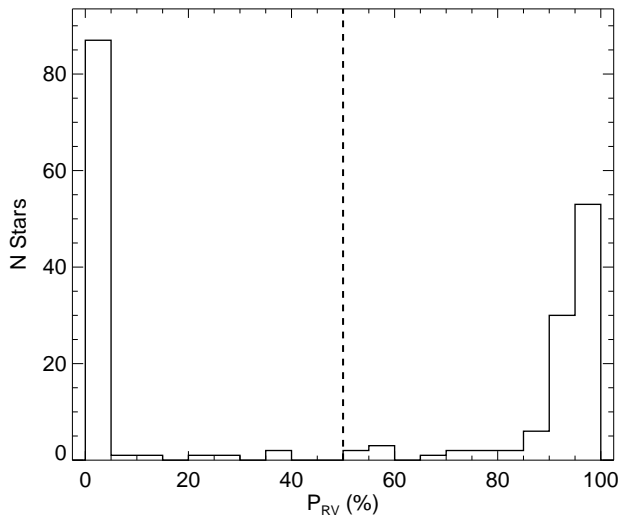}
\caption{RV membership probability histogram for single stars. We set an RV
  membership cutoff at 50\%, shown by the vertical dashed line.}
    \label{fig:probhist}
\end{figure}

Figure \ref{fig:probhist} presents a histogram of our RV membership
probabilities for single stars. As noted above, a clear separation is found
between cluster members and field stars. In keeping with previous WOCS
practice, we set our cutoff for RV membership at $P_{\rm RV}=50\%$,
designated by the vertical dashed line.

With our membership scheme established, we use the field distribution
presented in Figure \ref{fig:rvhist} to estimate the number of field-star
contaminants expected among our RV members. By integrating the model field
distribution over RVs producing $P_{\rm RV} \ge 50\%$, we expect 7 of our RV
members to be field stars.

From our RV membership determination, we define a star's membership ``Class'',
given in the penultimate column of Table \ref{tab:summary}, following
\citet{Gelleretal2008}. As presented in Section \ref{obs}, MC simulations show
that three RV measurements over the course of 200 days can detect, to
$\sim$90\% confidence, binaries with orbital periods up to 1000 days. This
baseline in time and number of observations is required for membership
determination and as such, any star not meeting this criterion is given an
unknown (U) classification in Table \ref{tab:summary}. A single star ($e/i \le
4$) meeting these conditions is designated a single member (SM) if $P_{\rm RV}
\ge 50\%$ or a single non-member (SN) if its RV probability falls below this
criterion. A velocity-variable star ($e/i > 4$) can fall into three classes:
(1) binary likely member (BLM) if the membership probability based on its mean
RV ($P_{\overline{\rm RV}}$) exceeds our criterion, (2) binary unknown (BU) if
$P_{\overline{\rm RV}}$ falls below 50\% but the range of measured RV crosses
the cluster mean, or (3) binary likely non-member (BLN) if the range in
measured RV does not cross the cluster mean. Once orbital solutions are found
for binary systems, $P_{\rm RV}$ will be computed based on the center-of-mass
velocity. At that point, binary stars will be classified as binary members
(BM) or binary non-members (BN) with the same criteria as single stars. Table
\ref{tab:class} lists the number of stars falling within each membership
class.

\subsection{Comparison of Proper-Motion and Radial-Velocity 
Memberships}
\label{ppmprv}

\begin{figure}[!t]
\centering
\includegraphics[keepaspectratio=true,scale=0.52]{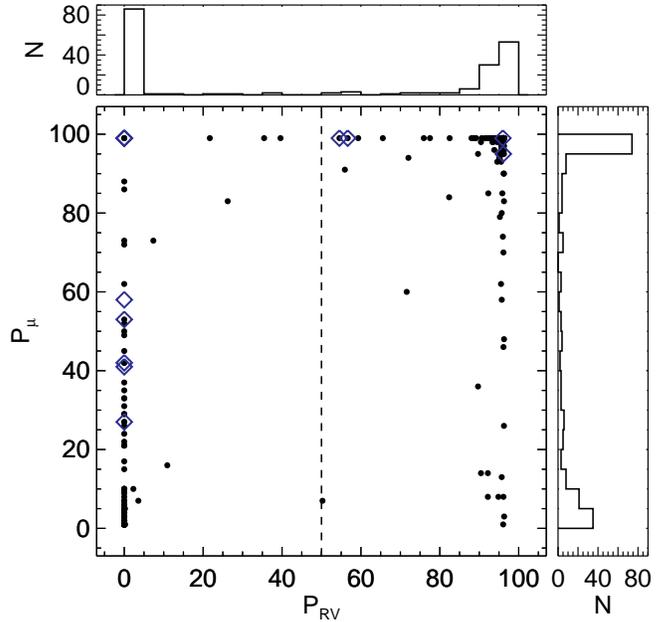}
\caption{A comparison of the P11 $P_\mu$ values (y-axis) with our $P_{\rm RV}$
  values (x-axis) for single stars. The dashed vertical line represents the
  radial-velocity membership criterion (50\%). The blue-HB candidates from P11
  are highlighted in blue diamonds. Blue diamonds without a central point are
  velocity variable. Their radial-velocity membership position reflects the
  membership probability of their mean radial velocity. Marginal distributions
  along each axis are presented in the top and right panels.}
    \label{fig:pmvrv}
\end{figure}

Of the 280 stars in our sample, 193 are single (non-velocity variable) stars
with RV membership determinations. Figure \ref{fig:pmvrv} presents the
comparison of RV and PM membership probabilities for all single stars. The
vertical dashed line marks the RV (50\%) membership criterion. All blue-HB
candidates from P11 are presented as blue diamonds. Note that some blue-HB
candidates are velocity variables and thus are presented as a blue diamond
without central points. Their RV membership positions represent the
probabilities of their mean RV ($P_{\overline{\rm RV}}$) and are subject to
change until an orbital solution is found (see Section \ref{HBCan} for full
details). In Figure \ref{fig:pmvrv}, we also present the marginal distributions
along each axis in the top and right panels. 

\begin{deluxetable}{ccccc}
\tablewidth{100pt}
\tabletypesize{\small}
\tablecaption{Number of Stars Within Each Membership Class}
\tablehead{
  \colhead{} &
  \colhead{Class} &
  \colhead{} &
  \colhead{N$_{\rm stars}$} &
  \colhead{}
}
\startdata
 & SM & & 101 & \\
 & SN & & 92 & \\
 & BLM & & 10 & \\
 & BU & & 6 & \\
 & BLN & & 12 & \\
 & U & & 59 & \\
\enddata
\label{tab:class}
\end{deluxetable}

Integrating the PM membership probabilities of stars with RV membership
determinations (193), $106\pm3$ cluster members are predicted, in good
agreement with the 101 RV members found. Breaking the stars with RV membership
determinations into three subsamples based on their PM membership
probabilities we find:

\begin{itemize}
\item{$86\pm1$ cluster members out of 89 are predicted in the $P_\mu > 80\%$
    sample while 81 RV members are found,}
\item{$17\pm2$ cluster members out of 37 are predicted from the $19\% \le
    P_\mu \le 80\%$ sample while 11 RV members are found,}
\item{$3\pm1$ cluster members out of 67 are predicted from the $1\% \le P_\mu
    < 19\%$ sample while 9 RV members are found.}
\end{itemize} 

\

Generally, we find fewer RV cluster members than the PM membership
probabilities would suggest. However, for stars with $1\% \le P_\mu < 19\%$, we
find $\sim$6 more RV members than the PM prediction. Recalling that 7 field
stars are expected among our RV members (Section \ref{membership}), it is
reasonable to suspect that some of our field contaminants are likely to be
drawn from the 9 RV members with $1\% \le P_\mu < 19\%$. As such, we note them
in our cleaned CMD (gray points; Figure \ref{fig:zcmd}(b)).

However, of these 9 low-$P_\mu$ RV members 7 fall on the cluster RGB or RC in
the cleaned CMD. Based on the CMD distribution of the 67 stars with $1\% \le
P_\mu < 19\%$ and RV membership determinations, a random draw of 9 such stars
would predict only one to fall in the RGB/RC region. For this reason we
suspect that most of these 9 stars are in fact cluster members and include
them with equal significance in our analyses below.

\

\section{DISCUSSION}
\label{disc}

\subsection{Color-Magnitude Diagram}
\label{cmd}

\begin{figure}[!tbh]
\centering
\includegraphics[keepaspectratio=true,scale=0.45]{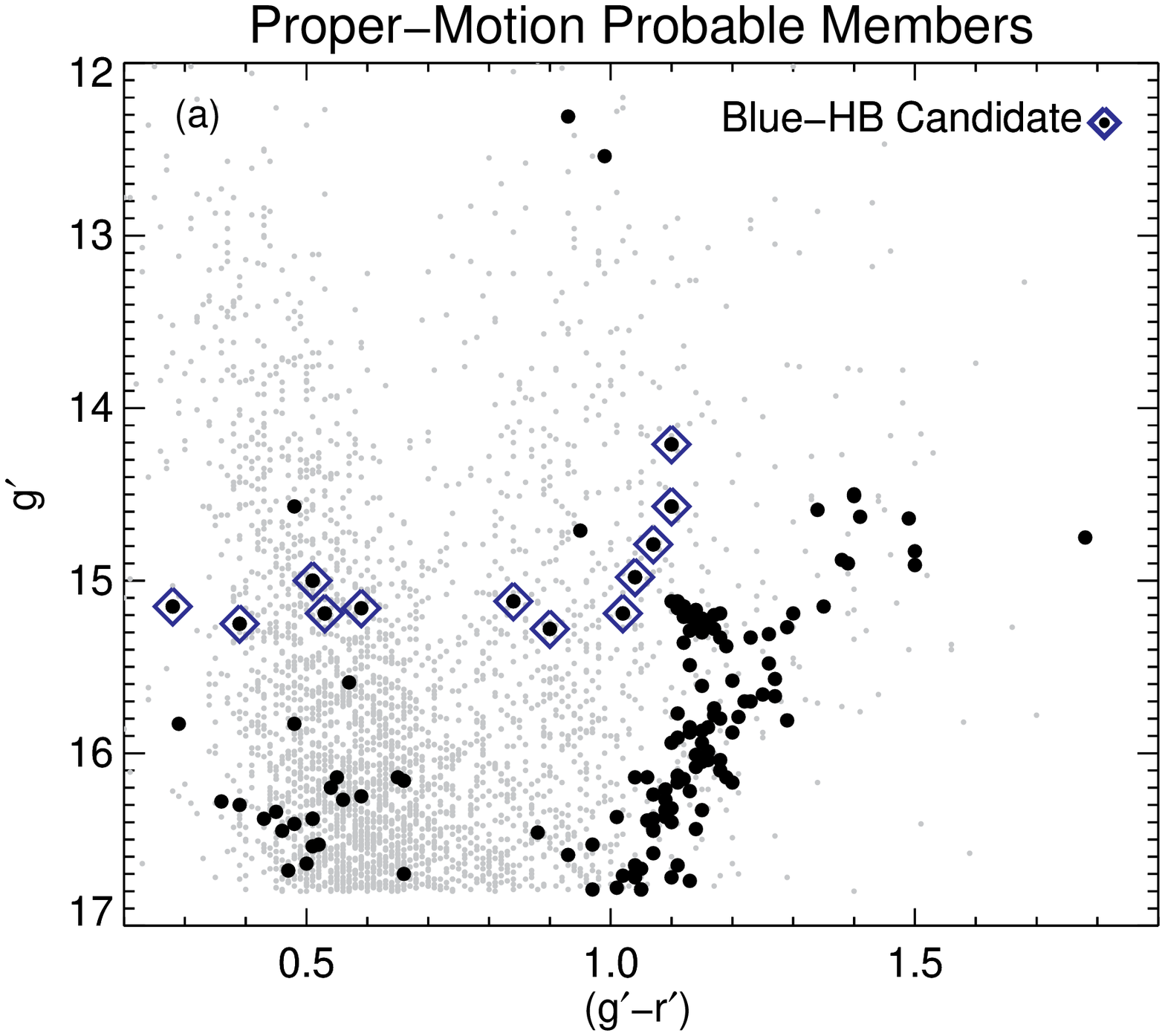}
\includegraphics[keepaspectratio=true,scale=0.45]{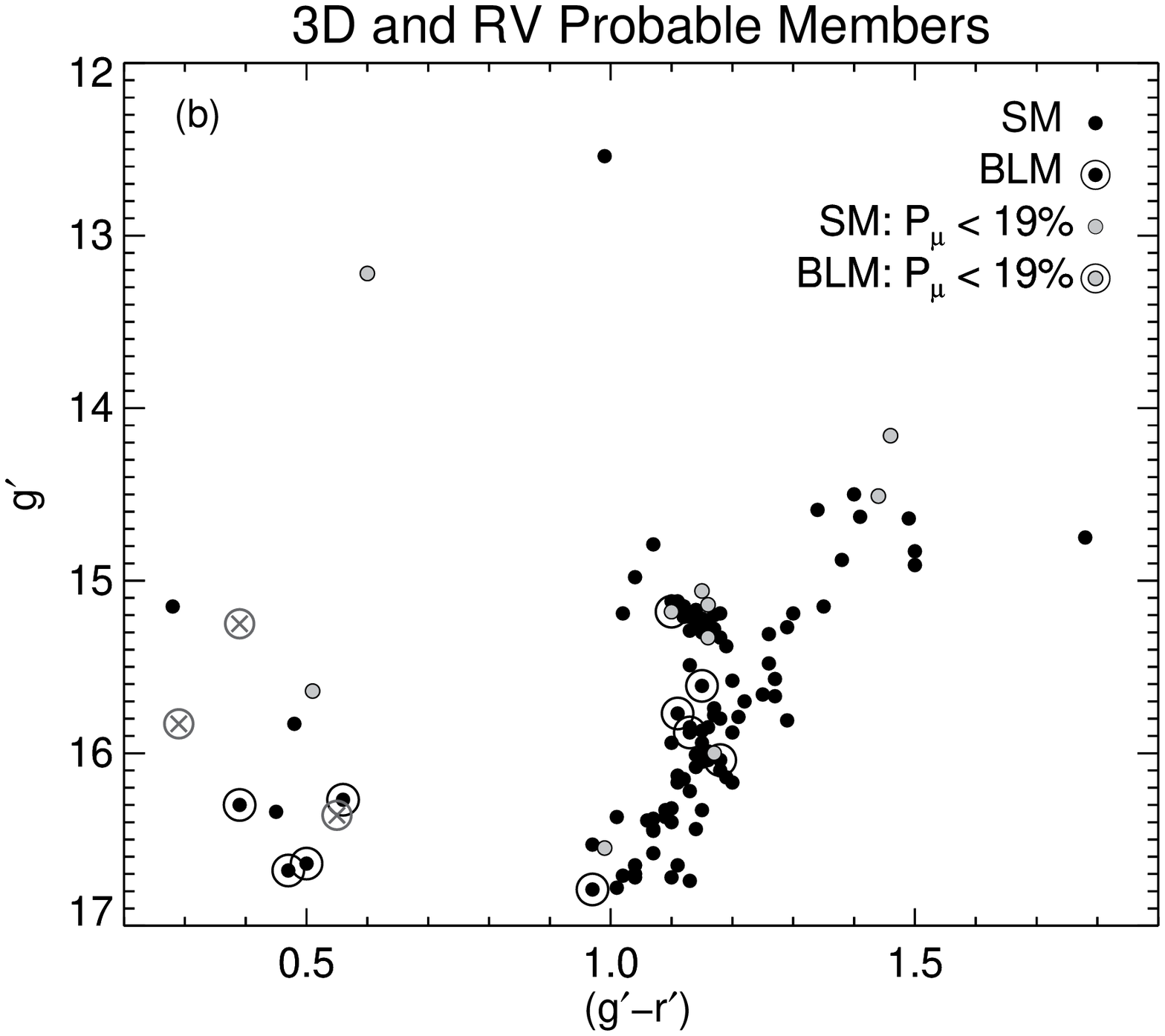}
\caption{(a) Upper CMD of proper-motion probable members ($P_\mu \ge 19\%$)
  from P11 with blue-HB candidates highlighted in blue diamonds. Small
  light-gray points are stars with $P_\mu < 19\%$. (b) CMD including
  radial-velocity membership determinations. Single members (SMs) are shown as
  black points, circled points indicate binary likely members (BLMs). Gray
  points are SMs and BLMs with $1\% \le P_\mu < 19\%$. Gray ``$\otimes$''
  symbols mark candidate binary members currently classified as binary
  unknowns (BUs) or binary likely non-members (BLNs), discussed in Section
  \ref{cmd}.}
    \label{fig:zcmd}
\end{figure}

Three-dimensional kinematic membership determinations using both PM and RV
measurements allow us to construct a cleaned CMD for the evolved population of
NGC 6791. Figure \ref{fig:zcmd}(a) displays the PM probable members ($P_\mu
\ge 19\%$) within our magnitude limit as black circles, with the 12 blue-HB
candidates marked in blue diamonds. Stars with $P_\mu < 19\%$ are shown as
light-gray points. The addition of RV membership information produces Figure
\ref{fig:zcmd}(b) and represents the most secure, systematic membership
determination to date for this cluster. In this lower panel, single members
(SMs) are marked as black points. Circled points signifying binary
likely members (BLMs). 

\

\

In Figure \ref{fig:zcmd}(b), we also include 3 candidate binary members falling
near the BSS population as gray ``$\otimes$'' symbols. Although these stars
are currently listed as BUs or BLNs in Table \ref{tab:summary}, based on their
RV measurements to date, we feel they have a high likelihood of becoming
members once orbital solutions are found. The two BUs (WOCS 54008 and 62041)
have large RV variations and few observations, which we suspect are
short-period binaries whose mean RV bears little significance on their RV
memberships. The other is a BLN (WOCS 46008) whose mean RV is within
3$\sigma$ of the cluster mean velocity. Due to the astrophysical significance
of binary stars to the formation of BSSs (see Section \ref{bs}), we note them
here for completeness but do not include them in the analysis below.

The lower panel of Figure \ref{fig:zcmd} reveals a richly populated RGB and RC
with 97 SM and BLM stars. We find an RG/RC binary fraction of 6.2\% for orbits
up to 1000 days, assuming that all BLMs are indeed members (fraction not
corrected for detection incompleteness). This binary fraction is less than the
global RG/RC binary fraction due to our insensitivity to long-period binaries.
Additionally, it is likely not representative of the main sequence binary
population given that the increased radii of post-main sequence stars will
alter or envelop the orbits of close binary companions. (We have not included
the 3 blue-HB candidates found just to the blue of the RC in the above
analysis, but we comment on them in Section \ref{HBCan}.)

The width of the RGB is roughly $(g^\prime$-$r^\prime) \sim 0.2$ mag, broader
than expected for a single-age stellar population. \citet{Brogaardetal2012}
derived a differential reddening map that, when applied on a star-by-star
basis, significantly narrowed the RGB.

With the RG/RC and BSS populations well defined, we empirically classify BSSs
within our sample as members falling blue of $(g^\prime$-$r^\prime)=0.7$. We
designate 8 SM and BLM stars as BSSs and find 4 to be velocity-variable. The
BSS population is discussed further in Section \ref{bs}.

Three stars in our CMD are assign to neither the RG/RC nor BSS population. They
are the two brightest, WOCS 57012 (SM; $P_{\rm RV}=55\%$, $P_\mu=91\%$) and
58012 (RV member only; $P_{\rm RV}=92\%$, $P_\mu=14\%$), and the reddest, WOCS
9018 (SM; $P_{\rm RV}=95\%$, $P_\mu=79\%$), stars on the cleaned CMD (Figure
\ref{fig:zcmd}(b)). Their CMD locations are not predicted by any phase of
stellar evolution at the age of NGC 6791, nor do we know of similarly located
members in other open clusters. Even given our typically high RV membership
probabilities, field contamination cannot be ruled out. Indeed, as noted
earlier, our membership probability model (Figure \ref{fig:rvhist}) predicts 7
field contaminants among the RV members.

\begin{deluxetable*}{rccccccrccc}
\tablewidth{0pt}
\tabletypesize{\footnotesize}
\tablecaption{NGC 6791 Blue-HB Proper-Motion Candidate Members}
\tablehead{
  \colhead{ID$_{\rm W}$} & 
  \colhead{ID$_{\rm P11}$} &
  \colhead{R.A.} &
  \colhead{Dec.} &
  \colhead{$g^\prime$} & 
  \colhead{$g^\prime$-$r^\prime$} &
  \colhead{$P_{\mu}$} & 
  \colhead{$\overline{\rm{RV}}$} &
  \colhead{$\sigma_{\rm{RV}}$} & 
  \colhead{$P_{\rm RV}$\,\tablenotemark{a}} & 
  \colhead{Comment\,\tablenotemark{b}} \\ 
  \colhead{} &
  \colhead{} &
  \colhead{} &
  \colhead{} &
  \colhead{} &
  \colhead{} &
  \colhead{} &
  \colhead{(km s$^{-1}$)} &
  \colhead{(km s$^{-1}$)} &
  \colhead{} &
  \colhead{}
}
\startdata
23004 & 69976 & 19:20:46.00 & 37:47:54.8 & 14.98 & 1.04 & 95 & -47.04 & 0.34 & 96 & 2-45\\
2001 & 65895 & 19:20:51.37 & 37:46:30.5 & 14.79 & 1.07 & 99 & -47.86 & 0.34 & 95 & NW-20\\
15007 & 60456 & 19:21:07.20 & 37:44:34.7 & 15.19 & 1.02 & 99 & -44.80 & 0.51 & 56 & 3-27\\
4003 & 64589 & 19:20:59.95 & 37:46:03.3 & 15.15 & 0.28 & 99 & -44.76 & 0.62 & 54 & 2-17\\
46008&64067& 19:20:34.98 & 37:45:52.3 & 15.25 & 0.39 & 99 & -40.89 & 5.29 &(0)& S2746; RR; SB1\\
7021&89107 & 19:20:33.03 & 37:55:55.9 & 14.21 & 1.10 & 41 & -75.11 & 4.18 & (0) & S2382; SB1 \\
8031 & 93625 & 19:21:42.32 & 37:58:05.5 & 15.16 & 0.59 & 58 & -18.89 & 4.25 & (0) & SB1\\
18007 & 58857 & 19:21:02.55 & 37:43:59.5 & 15.19 & 0.53 & 99 &  15.57 & 0.81 & 0 & 3-22\\
9007 & 69734 & 19:21:08.07 & 37:47:49.4 & 14.57 & 1.10 & 99 & -22.84 & 0.33 &  0 & 3-33\\
21016 & 82982 & 19:21:14.70 & 37:53:07.5 & 15.28 & 0.90 & 53 &  11.47 & 0.29 &  0 & S14140\\
19018 & 52460 & 19:20:17.44 & 37:41:23.9 & 15.12 & 0.84 & 42 &  10.05 & 0.29 &  0 & S440\\
11019 & 45106 & 19:20:31.40 & 37:38:09.7 & 15.00 & 0.51 & 27 & -54.49 & 0.28 &  0 & S2101 \\
\enddata
\tablecomments{Photometry, coordinates, and proper-motion membership
  probabilities come from I. Platais et al (2014, in preparation).}
\tablenotetext{a}{RV membership probabilities in parenthesis indicate the
  probability of the mean RV for velocity variables ($P_{\overline{\rm RV}}$)
  and remains uncertain until a binary orbital solution is found.}
\tablenotetext{b}{\citet{Stetsonetal2003} IDs are preceded by an ``S'',
  remaining IDs are from \citet{Kinman1965}. SB1 marks single lines
  spectroscopic binaries, RR marks rapid rotators.}
\label{tab:hbcan}
\end{deluxetable*}

\subsection{Blue-Horizontal Branch Candidates}
\label{HBCan}

Table \ref{tab:hbcan} summarizes our RV membership results for the 12 blue-HB
candidates from P11. Here we provide the WOCS IDs (ID$_{\rm W}$) with a cross
reference to the ID given in P11 (ID$_{\rm P11}$) in the first two
columns. We find four of these stars we find to be SMs. Three (WOCS 2001, 23004,
15007) fall just to the blue of the RC (see CMD in Figure \ref{fig:zcmd}(b))
and the fourth, WOCS 4003, is the bluest star in the cleaned CMD, lying in the
BSS regime.

Of the remaining 8 blue-HB candidates, 5 are single stars that are RV
non-members ($P_{\rm RV}=0$). Two, while velocity variable (BLN), have mean
velocities far from the cluster mean; with a minimum of 12 RV measurements for
each, we conclude that they are also RV non-members. The last blue-HB candidate,
WOCS 46008, is a velocity variable, rapid rotator listed as a BLN in Table
\ref{tab:summary}. With a mean RV 2.7$\sigma$ (6.5 km s$^{-1}$) from the
cluster mean velocity we consider it a potential binary member and include it
in Figure \ref{fig:zcmd} as the brightest gray ``$\otimes$'' symbol in the BSS
region of the CMD. An orbital solution will be required to securely determine
this star's RV membership.

The results of our RV membership study do not support the presence of a
continuous blue-HB population suggested by P11. Instead, we find 1 SM to
reside with BSSs in the CMD and 3 SMs that fall just blue of the RC. This
observational finding is consistent with the \citet{Brogaardetal2012}
theoretical conclusion that a blue-HB, if one were to exist, would be more
luminous than the P11 proposed candidates.

Even so, the three SMs falling near, but to the blue of, the RC remain
puzzling as they cannot be explained by the typical photometric error
($\sim$0.02 mag). Although our field-star and cluster population model
predicts 7 field contaminants in Figure \ref{fig:zcmd}(b), these 3 stars have
high PM and RV membership probabilities (Table \ref{tab:hbcan}), leaving a
small chance that all 3 are field stars.

We speculate that these 3 may be evolved descendants of BSSs. If, for
simplicity, we model these stars with normal stellar evolutionary tracks, we
find that all three lie on the giant branch of a 1.3 Gyr isochrone, implying
masses of $\sim 2$ $M_\odot$. (Here we use a reddening of
E$(g^\prime$-$r^\prime)=0.165$, an extinction of $A_{g^\prime}=0.65$ mag, an
age of 8 Gyr, [Fe/H]=+0.30 \citep{Boesgaardetal2009}, and a distance of 3.8
kpc, obtained by matching Padova isochrones \citep{Bressanetal2012} to the NGC
6791 RC and RGB.) This stellar mass is roughly 75\% more massive than a
typical NGC 6791 giant \citep[1.2 $M_\odot$;][]{Basuetal2011}. Such a mass is
not unprecedented for a BSS, which in this case would have formed $\sim$ 1 Gyr
ago. We note that it is surprising that all 3 stars do not have detected
velocity variability given the high BSS binary fraction found in NGC 6791
(Section \ref{bs}) and in other open clusters \citep{Geller&Mathieu2012}. This
may point to stellar collisions or mergers as the formation mechanism of our
evolved BSSs candidates \citep{Leigh&Sills2011}.

The SM status of our fourth blue-HB candidate, WOCS 4003 (2-17;
\citealt{Kinman1965}), confirms the membership first proposed by
\citet{Greenetal1996} as a blue-HB star. A subsequent analysis of this star by
\citet{Brogaardetal2012}, however, found its spectral properties to be most
consistent with a mass of 1.9 $M_\odot$, whereas the expected mass of an HB
star is $\sim0.6 M_\odot$. The large mass along with low luminosity (for a
blue-HB star) led these authors to conclude a BSS classification.

Again assuming standard stellar evolutionary tracks, we find this star to lie
on a $\sim$0.9 Gyr isochrone with a mass of $\sim2 M_\odot$. We conclude (in
agreement with \citealt{Brogaardetal2012}) that WOCS 4003 is a BSS member
that formed perhaps $\sim1$ Gyr ago. Given the large mass and non-varying RV,
this BSS may also be a candidate for formation via a merger or collision.

\

\subsection{Blue Stragglers}
\label{bs}

Table \ref{tab:bs} presents the membership information for BSS members and
candidate members. We break our BSS sample into two groups based on their
membership likelihood. First, as ordered in Table \ref{tab:bs}, are SMs and
BLMs, totaling 8 stars. The second is comprised of 3 potential BSS members,
all velocity variables (BU and BLN), whose large amplitude of variability
($e/i > 4$) and small number of RV measurements make their RV memberships
uncertain. We suspect these stars may be cluster members. They are plotted as
gray ``$\otimes$'' symbols in the BSS region of Figure \ref{fig:zcmd}(b).

\begin{deluxetable*}{ccccccrrcc}
\tablewidth{0pt}
\tabletypesize{\footnotesize}
\tablecolumns{10}
\tablecaption{NGC 6791 Blue Straggler Membership}
\tablehead{
  \colhead{ID$_{\rm W}$} & 
  \colhead{R.A.} &
  \colhead{Dec.} &
  \colhead{$g^\prime$} & 
  \colhead{$g^\prime$-$r^\prime$} &
  \colhead{$P_{\mu}$} & 
  \colhead{$\overline{\rm{RV}}$} &
  \colhead{$\sigma_{\rm{RV}}$} & 
  \colhead{$P_{\rm RV}$\,\tablenotemark{a}} & 
  \colhead{Comment\,\tablenotemark{b}} \\ 
  \colhead{} &
  \colhead{} &
  \colhead{} &
  \colhead{} &
  \colhead{} &
  \colhead{} &
  \colhead{(km s$^{-1}$)} &
  \colhead{(km s$^{-1}$)} &
  \colhead{} &
  \colhead{}
}
\startdata
\sidehead{Single Members \& Binary Likely Members}
\hline
4003 & 19:20:59.95 & 37:46:03.3 & 15.15 & 0.28 & 99 & -44.76 & 0.62 & 54 & Blue-HB Can. \\
21028 & 19:21:56.64 & 37:52:05.8 & 15.64 & 0.51 & 8 & -46.06 & 1.04 & 92 &  \\
33025 & 19:19:52.07 & 37:48:31.6 & 16.34 & 0.45 & 60 & -45.10 & 0.97 & 71 &  \\
70011 & 19:21:12.68 & 37:49:48.7 & 15.83 & 0.48 & 97 & -47.02 & 0.37 & 96 &  \\
11003 & 19:20:58.31 & 37:47:08.1 & 16.64 & 0.50 & 99 & -46.07 & 1.59 & (92) & SB1 \\
22004 & 19:21:01.05 & 37:47:34.7 & 16.68 & 0.47 & 99 & -48.21 & 3.77 & (95) & SB1 \\
57009 & 19:20:53.91 & 37:42:26.0 & 16.27 & 0.56 & 99 & -49.62 & 2.11 & (75) & SB1 \\
58008 & 19:20:10.42 & 37:47:42.1 & 16.30 & 0.39 & 99 & -47.76 & 5.08 & (96) & SB1, RR \\
\hline
\sidehead{Candidates}
\hline
46008&19:20:34.98&37:45:52.3 & 15.25 & 0.39 & 99 & -40.05 & 4.39 & (0) & SB1, RR, Blue-HB Can.\\
54008 & 19:21:10.71 & 37:45:31.4 & 15.83 & 0.29 & 99 & -55.34 & 10.99 & (0) & SB1, RR \\
62041 & 19:21:57.88 & 37:30:56.9 & 16.36 & 0.55 & 2 & -53.01 & 28.80 & (0) & SB1 \\
\enddata
\tablecomments{Photometry, coordinates, and proper-motion membership
  probabilities come from I. Platais et al (2014, in preparation).}
\tablenotetext{a}{RV membership probabilities in parenthesis indicate the
  probability of the mean RV for velocity variables ($P_{\overline{\rm RV}}$)
  and remains uncertain until a binary orbital solution is found.}
\tablenotetext{b}{SB1 marks single lines spectroscopic binaries, RR marks
  rapid rotators.}
\label{tab:bs}
\end{deluxetable*}

We note that 1 BLM BSS (WOCS 58008) and 2 binary BSS candidates (WOCS 46008,
54008) are designated as rapid rotators. While orbital solutions are still
required to confirm the membership of these stars, the distribution of BSS
rotation rates may prove to be a valuable test of BSS formation theories
\citep{Sillsetal2005}.

Observations to date yield an NGC 6791 BSS population of 8 stars, 4 being SMs
and 4 being BLMs. Among the 8 BSS SMs and BLMs, we find a binary fraction of
50\% for binaries with orbits up to 1000 days, assuming that all the BLMs are
indeed members. The three candidate BSSs would only increase this binary
frequency. This is higher than both the RG/RC binary frequency found above and
the main sequence binary fraction determined by the \citet{Janes&Kassis1997}
analysis of the CMD binary main sequence (14\%). Although still preliminary, the
high binary fraction of NGC 6791 BSSs agrees with the findings of
\citet{Geller&Mathieu2012} for NGC 188 BSSs ($76\pm19\%$).  This suggests that
binaries may play a key role in the formation of BSSs in NGC 6791. A complete
analysis of the binary BSS orbital solutions is required before determining
the roles of any specific BSS formation mechanism.

\subsection{Dynamical Mass}
\label{mass}

Our systematic RV survey finds $\sigma_{6791}=1.1\pm0.1$ km s$^{-1}$ from
our Gaussian fitting procedure (Section \ref{membership}). However, undetected
binaries in our sample broaden the wings of the RV distribution and inflate
our observed dispersion value. Thus, this Gaussian-fit RV dispersion is an
upper limit.

\begin{figure}[!tbh]
\centering
\includegraphics[keepaspectratio=true,scale=0.45]{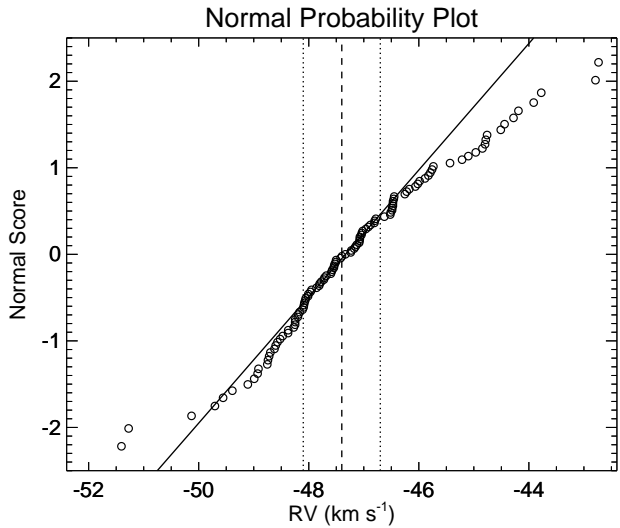}
\caption{Normal probability plot of the mean radial-velocity for single
  stars. The vertical dashed line marks the cluster mean velocity from Figure
  \ref{fig:rvhist}. The solid line is a linear best-fit to the data within the
  central 1.4 km s$^{-1}$ represented by vertical dotted lines. The inverse
  slope of this line is a measure of the observed cluster radial-velocity
  dispersion, yielding $0.73\pm0.09$ km s$^{-1}$.}
    \label{fig:npp}
\end{figure}

Modeling the effects of binarity on the observed RV dispersion is beyond the
scope of this paper. Instead, we use the normal probability plot
\citep{Chambers1983} to derive a measure of the observed RV dispersion that is
less sensitive to the wings of the distribution. The normal probability plot
in Figure \ref{fig:npp} compares our mean RV measurements for single stars
(open circles) to a normal distribution. Measurements drawn from a normal
distribution map into a straight line whose inverse slope corresponds to the
standard deviation of the parent distribution, which in this case is the
observed RV dispersion (before correction for measurement errors).

Using the same RV center from our Gaussian fit (Section \ref{membership}),
denoted by the dashed vertical line, we fit a line to the central 1.4 km
s$^{-1}$ (vertical dotted lines). We justify fitting to only this central
region because this range of data is normally distributed. Fits to more
centralized regions reproduce the same result with larger error. Outside this
range, it is clear the RV measurements deviate from Gaussian, possibly from
binary contamination. Our best-fit line yields an observed RV dispersion of
$0.73\pm0.09$ km s$^{-1}$. We then deconvolve the best fit RV dispersion with
our measurement precision (0.38 km s$^{-1}$) to obtain a true one-dimensional
velocity dispersion of $\sigma_c=0.62\pm0.10$ km s$^{-1}$.

We calculate the cluster mass using the equation for dynamical mass in
projection from \citet{Spitzer1987},
\begin{equation}
    M=\frac{10 \langle \sigma_c^2 \rangle R_h}{G},
\label{eqn:dmass}
\end{equation}
where $\sigma_c$ is the one-dimensional velocity dispersion of the cluster and
$R_h$ is the projected half-mass radius. Equation \ref{eqn:dmass} assumes a
gravitational potential energy prefactor of 0.4, the observed (projected)
half-mass radius measured, $R_h$, is $3/4$ of the true half-mass radius, and
$\sqrt{3}\sigma_c=\sigma_{\rm 3D}$ \citep{Spitzer1987}.

Using $R_h=5.1\pm0.2$ pc from King model fitting (P11), we find a mass of
${4600}\pm{1500}$ $M_\odot$. Current literature values for the mass of NGC
6791 based on photometry range from $\ge4000$ $M_\odot$
\citep{Kaluzny&Udalski1992} to $\sim$5000 $M_\odot$ (P11), in good agreement
with our dynamical mass determination. In the context of other open clusters,
our derived mass places NGC 6791 in the top 5\% of the
\citet{Piskunovetal2008} open cluster sample, complete to 850 pc.

\subsection{Stars of Note}
\label{son}

\subsubsection{WOCS 54008} 
WOCS star 54008 is currently listed as a rapidly rotating ($v$sin $i=67.2$ km
s$^{-1}$) SB1 with unknown membership (BU) and a $P_\mu=99\%$. In Figure
\ref{fig:zcmd} it is the bluest candidate binary member shown with a gray
``$\otimes$'' symbol. \citet{Mochejskaetal2005} find this source (V106) to be
an eclipsing W UMa type contact system with a period of 1.4464 days. In our
CMD, we would consider WOCS 54008 to be a BSS if its membership is confirmed.

W UMa BSSs have been found in open clusters (M 67, S1036;
\citealt{Sandquist&Shetrone2003}) and globular clusters (M 30, 12005407B;
\citealt{Lovisietal2013}), pointing to mass transfer as a mechanism for the
formation of at least some BSSs. While W UMa-type systems have been found with
both decreasing \citep{Qianetal2013} and increasing
\citep{Christopoulou&Papageorgiou2013} orbital periods, either the eventual
coalescence or detached binary result is predicted to form a BSSs
\citep{Qianetal2006}.

\subsubsection{WOCS 43033}
An SB2 categorized as a BU, WOCS 43033 has RV measurements that cross the
cluster mean velocity and a $P_\mu=6\%$. Given its membership class and CMD
location ($g^\prime = 15.82$; $g^\prime$-$r^\prime = 0.55$), WOCS 43033 would
normally be placed in the candidate BSS section. Several observations of this
system were obtained at phases when the primary and secondary CCF peaks were
superimposed. Observations of superimposed peaks reveals an estimate of the
center-of-mass velocity of the system, measured to be $\sim-27$ km s$^{-1}$.
Although this method is not as precise as measuring the center-of-mass
velocity from a full orbital solution, the resulting RV is sufficiently far
from the cluster mean of $-47.40$ km s$^{-1}$ that cluster membership is
unlikely.  Therefore, WOCS 43033 is neither listed in Table \ref{tab:bs} as a
candidate BSS nor shown in Figure \ref{fig:zcmd}(b) with a gray ``$\otimes$''
symbol.  Nonetheless, observations will continue in order to obtain a full
orbital solution and definitively determine its RV membership probability.

\subsubsection{WOCS 54034}
Similar to WOCS 43033 above, WOCS 54034 is an SB2 that falls in the BSS region
of the CMD. It has a $P_\mu=33\%$ and is classified as a BU from our RV
measurements. We determine the center-of-mass velocity of this system to be
$\sim-32$ km s$^{-1}$ from observations with superimposed primary and
secondary CCF peaks. Due to its large separation from the cluster mean
velocity, we find WOCS 54034's RV membership unlikely and, for this reason, we
exclude it from Figure \ref{fig:zcmd} and Table \ref{tab:bs}. Observations of
this star will continue until an orbital solution is found.

\subsubsection{Kepler Selected Targets}
\citet{Stelloetal2011}, using Fourier analysis of {\sl Kepler} lightcurves,
determined an asteroseismic membership criterion based on the average large
frequency separation, $\Delta\nu$, and frequency of maximum oscillation power,
$\nu_{\rm max}$. These authors find 65 seismic RG/RC members and 3 probable or
uncertain members. A cross reference between our catalogs finds 100\%
agreement between the three-dimensional kinematic and seismic membership
results for the 48 sources (2 are BLMs) which have RV membership
determinations. In addition, we find KIC 2437171 (WOCS 3006), which is blended
on the {\sl Kepler} CCD, and potential member KIC 2438421 (WOCS 43010) to be
SMs.

Continued {\sl Kepler} monitoring of the \citet{Stelloetal2011} vetted targets
by \citet{Corsaroetal2012} established stellar mass and radius
estimates as well as distinguished between RG (H-shell burning) and RC (He-core
burning) stars. Four stars from this analysis were deemed outliers based on
inconsistencies between their determined mass and/or energy production
mechanism (H-shell/He-core) and their CMD locations. We comment on the
membership on these stars below:

KIC 2437589 (WOCS 16007). This star is classified as an SM with
$P_\mu=99\%$ and $P_{\rm RV}=91\%$. On the CMD, this star falls in the RC but
has the asteroseismic properties of a massive 1.7 $M_\odot$ RG, whereas the
mass of typical NGC 6791 RGB stars is 1.2 $M_\odot$
\citep{Basuetal2011}. \citet{Brogaardetal2012} has suggested that this star is
an evolved BSS in the RG phase.

KIC 2436417 (WOCS 41008). A SM with $P_\mu=99\%$ and 
$P_{\rm RV}=95\%$, this star has the same mass and asteroseismic properties as
other RC stars, but with a larger radius. This is a proposed ``evolved RC
star'' on its way to the asymptotic giant branch. The larger radius is also
consistent with the P11 CMD location as one of the brightest RC stars.

KIC 2437804 (WOCS 4004). Also a proposed evolved RC star, this star
has $P_\mu=99\%$ and $P_{\rm RV}=89\%$. This star is also one of the brightest RC
stars.

KIC 2437103 (WOCS 3003). With only 2 RV measurements, this star is
labeled as unknown (U) in Table \ref{tab:summary}. The mean of its two RV
measurements ($-49.0$ and $-49.2$ km s$^{-1}$) meet the RV membership
criterion ($P_{\overline{\rm RV}}=87\%$) and it has a $P_\mu=10\%$. This star
is of note because it falls on the RGB with the asteroseismic properties of an
RC star.

\section{CONCLUSIONS}
\label{conc}

This program marks the first systematic, high-precision ($\sim$0.38 km
s$^{-1}$) RV survey of the old, metal-rich open cluster NGC 6791. Combining
our RV information with proper-motion memberships from I. Platais et
al. (2014, in preparation), we derive three-dimensional kinematic membership
probabilities for 193 stars. The resulting members define the evolved
population of the cluster. Here, we list the main findings of our work.

\begin{itemize}

\item We find 91 single cluster members and 6 binary likely members in the RGB
  and RC.

\item Of the 12 blue-HB candidates suggested by \citet{Plataisetal2011} we
  find 4 to be single cluster members. Three fall just blue of the RC, which
  we speculate may be evolved BSSs. The last we conclude to be a BSS. Evidence
  for a blue-HB population is not supported by our RV membership study.

\item Four single members and 4 binary likely members are classified as
  BSSs. Three additional binaries with large velocity variability (short
  period) may prove to be BSS cluster members once orbital solutions are
  found.

\item We find a 6.2\% RG binary frequency and a 50\% BSS binary frequency,
  assuming all binary likely members are indeed members.

\item The radial velocity of NGC 6791 is $-47.40\pm0.13$ km s$^{-1}$.

\item With normal probability analysis of our radial-velocity measurement
  distribution we find a cluster RV dispersion of
  $\sigma_c=0.62\pm0.10$ km s$^{-1}$. This corresponds to a
  dynamical mass of $4600\pm1500 M_\odot$.

\end{itemize}

Our RV survey of NGC 6791 from the WIYN telescope remains ongoing to complete
membership statistics, determine orbital solutions for binaries, and
investigate the anomalous cluster members of our cleaned CMD including BSSs
and potential evolved BSSs.

\

\acknowledgments The authors thank E. Leiner and K. Milliman for
many nights of observing and hours of data reduction for this project. We
also thank the referee for a thorough read of our paper and
helpful comments. Support for this program is provided by NSF grant
AST-0908082.

\

% Begowatts, Begowatts, Begowatts. Bega-what?!? ... The Begowatts. @thebegowatts #thebegowatts

\end{document}